\documentclass[]{proarticle}
\usepackage{multicol}
\usepackage{graphicx}
\usepackage{amssymb}
\usepackage{epsfig}

\def\bea{\begin{eqnarray}}
\def\eea{\end{eqnarray}}
\begin{document}
\title{BRST-BFV Lagrangian formulations for Higher-Spin fields  subject to two-column Young tableaux.}
\author{A.A. Reshetnyak}
\maketitle
\address{Department of Theoretical Physics, Tomsk State Pedagogical University, Kievskaya str., 60, 634061 Tomsk, Russia,\\
Institute of Strength Physics and Material Science
Siberian Branch of RAS,  Akademicheskii av.  2/4, 634021 Tomsk, Russia.}
\eads{reshet@ispms.tsc.ru, reshet@tspu.edu.ru}
\begin{abstract}
The details of  Lagrangian  description of irreducible integer
higher-spin representations of the Poincare group with an
Young tableaux $Y[\hat{s}_1,\hat{s}_2]$ having  $2$ columns  are considered for Bose particles propagated on an arbitrary dimensional Minkowski space-time. The procedure is based, first, on using of an auxiliary Fock space generated by Fermi oscillators (antisymmetric basis), second, on construction of the Verma module  and finding auxiliary oscillator realization for  $sl(2)\oplus sl(2)$ algebra which encodes the second-class operator constraints subsystem in the HS symmetry superalgebra. Application  of an  BRST-BFV receipt  permits to reproduce gauge-invariant Lagrangians with reducible gauge symmetries describing the free dynamics of both massless and massive mixed-antisymmetric bosonic fields of any spin with appropriate number of gauge and Stueckelberg fields. The general prescription possesses by the  possibility to derive constrained Lagrangians with only BRST-invariant extended algebraic constraints which describes the Poincare group irreducible representations in terms of mixed-antisymmetric tensor fields with 2 group indices.
\end{abstract}

%% \MSC 78A35 \sep 78-05 \sep 81V45

\begin{multicols}{2}
\section{Introduction}
%%%%%%%%%%%%%%%%%%%%%%%%%%%%%%%%%%%%%%
The belief to reconsider the problems of an unique
description of variety of elementary particles and  known
interactions maybe resolved within higher-spin (HS) field theory whose revealing together with the proof of
supersymmetry display,  and finding a new insight on origin of Dark Matter remains by the aims in LHC experiment programm
(\cite{LHC}). Because of the existence of so-called  tensionless limit in the (super)string theory \cite{tensionlessl}, which operates with an infinite tower of
 HS fields with integer and half-integer spins, the  HS field theory   may be considered both of superstring theory part and as an method
to study a superstring theory structure. On present state of HS field theory
we recommend to know from reviews \cite{reviews}, \cite{reviewsV}, \cite{reviews1},  \cite{reviews3}. The paper considers
the  results of constructing
 Lagrangian formulations (LFs) for free integer   both massless and
 massive mixed-antisymmetry tensor HS fields  on flat
$\mathbb{R}^{1,d-1}$-space-time subject to arbitrary Young
tableaux (YT) with $2$ columns $Y[\hat{s}_1,\hat{s}_2]$ for $\hat{s}_1 \geq \hat{s}_2 $  in Fronsdal metric-like formalism
on a base of BFV-BRST approach \cite{BFV}, and  precesses  the results
which appear soon in \cite{BurdikReshetnyak} (as well as for fermionic mixed-antisymmetric  spin-tensor HS fields on  $\mathbb{R}^{1,d-1}$-space-time subject to arbitrary  $Y[\hat{n}_1 + \frac{1}{2},\hat{n}_2 + \frac{1}{2}]$ in
\cite{BuchbinderTakReshetnyak}).

The irreducible Poincare
or (Anti)-de-Sitter ((A)dS) group representations in the constant curvature space-times may be described both by mixed-symmetric HS fields subject to arbitrary YT with $k$ rows, $Y({s}_1,...,{s}_k)$, (case of symmetric basis) determined by more than one spin-like parameters ${s}_i$ \cite{Labastida},  \cite{metsaevmixirrep}  and, equivalently, by mixed-antisymmetric tensor or spin-tensor fields subject to  arbitrary YT now with $l$ columns, $Y[\hat{s}_1,...,\hat{s}_l]$,  (case of antisymmetric basis) with integers or half-integers $\hat{s}_1\geq \hat{s}_2\geq ...\geq \hat{s}_l$ having a spin-like interpretation \cite{BurdikReshetnyak}, \cite{BuchbinderTakReshetnyak}.
   Both mixed-symmetric and mixed-antisymmetric HS fields appear  for  $d>4$ space-time dimensions,
in addition to totally symmetric and antisymmetric irreducible representations of Poincare
or (A)dS algebras. Whereas  for the latter
ones and for mixed-symmetric HS fields case the LFs both for massless and massive free higher-spin fields
is well enough developed \cite{massless Minkowski},  \cite{massless AdS},
\cite{massive AdS},  \cite{SkvortsovZinoviev}, \cite{Franciamix}, \cite{Zinovievta} as well as on base of BFV-BRST approach, e.g.
in \cite{BurdikPashnev}--\cite{Reshetnyk2}, for the mixed-antisymmetric case the problem
of their field-\-theoretic description has not yet solved except for the constrained bosonic fields subject to $Y[\hat{s}_1,\hat{s}_2]$ on the level of the equations of motion only \cite{Alkalaev} in so-called "frame-like"
 formulation.

We use, first, the conventions for the metric tensor $\eta_{\mu\nu} = diag (+,
-,...,-)$, with Lorentz indices $\mu, \nu = 0,1,...,d-1$, second,  the  notation $\epsilon(A)$, $gh(A)$ for
the respective values of Grassmann parity and ghost number of a
quantity $A$, and denote by $[A,\,B\}$ the supercommutator of
quantities $A, B$, which for theirs definite values of
Grassmann parities is given by $[A\,,B\}$ = $AB -
(-1)^{\epsilon(A)\epsilon(B)}BA$.

\section{Derivation of Integer
HS Symmetry Superlgebra on $\mathbb{R}^{1,d-1}$ }\label{sec2}
%
%%%%%%%%%%%%%%%%%%%%%%%%%%%%%%%%
We consider a massless integer spin irreducible
representation of Poincare group in a  Minkowski space
$\mathbb{R}^{1,d-1}$ which is described by a tensor field
$\Phi_{[\mu^1]_{\hat{s}_1},[\mu^2]_{\hat{s}_2}}
\hspace{-0.2em}\equiv \hspace{-0.2em}
\Phi_{\mu^1_1\ldots\mu^1_{\hat{s}_1},\mu^2_1\ldots\mu^2_{\hat{s}_2}}$
 of rank $\hat{s}_1+\hat{s}_2$ and generalized spin
 $\mathbf{s} \equiv (s_1,...,s_{s_2}; s_{s_2+1},...,s_{s_1})$ = $(2, 2,
 ... , 2; 1, ..., 1)$, (with omitting later a sign  "$\hat{\phantom{s}}$" under $\hat{s}_i$ and $s_1 \geq s_2 > 0, s_1
\leq [d/2])$ subject to a YT  with $2$ columns
of height  $s_1, s_2$, respectively
\begin{equation}\label{Young k}
\Phi_{[\mu^1]_{s_1},[\mu^2]_{s_2}}
\hspace{-0.3em}\longleftrightarrow \hspace{-0.3em}
\begin{array}{|c|c|}\hline%\vphantom{\biggm|}
  \!\mu^1_1 \!&\! \mu^2_1\!  \\
   \cline{1-2} \!\cdot\! &   \!\cdot\!   \\
   \cline{1-2}
    \! \mu^1_{s_2}\! &\! \mu^2_{s_2}\!    \\
  \cline{1-2} \!\cdots\!\\
  \cline{1-1} \!\mu^1_{s_1}\! \\
  \cline{1-1}
\end{array}\ ,
\end{equation}
This field is antisymmetric with respect to the permutations of each
type of Lorentz indices $\mu^i$,
  and
obeys to the Klein-Gordon (\ref{Eq-0b}), divergentless
(\ref{Eq-1b}), traceless (\ref{Eq-2b}) and mixed-antisymmetry
equations (\ref{Eq-3b}) :
\begin{eqnarray}
\label{Eq-0b} &&\hspace{-1.5em}
\partial^\mu\partial_\mu\Phi_{[\mu^1]_{s_1},[\mu^2]_{s_2}}
 =0,\\
&& \hspace{-1.5em} \partial^{\mu^i_{l_i}}\Phi_{
[\mu^1]_{s_1},[\mu^2]_{s_2}} =0, \texttt{ for } 1 \leq l_i \leq s_i,\ i=1,2,\label{Eq-1b}
\\
&& \hspace{-1.5em}   \eta^{\mu^1_{l_1}\mu^2_{l_2}}\Phi_{
[\mu^1]_{s_1},[\mu^2]_{s_2}}=0,\texttt{ for }
 1 \leq l_i \leq s_i,  \label{Eq-2b}\\
&& \hspace{-1.5em}\Phi_{
[[\mu^1]_{s_1},\underbrace{\mu^2_1...\mu^2_{l_2-1}}\mu^2_{l_2}]...\mu^2_{s_2}}=0, \label{Eq-3b}
\end{eqnarray}
  where the  bracket below in (\ref{Eq-3b}) denotes that the indices  in it are not
included into  antisymmetrization, i.e. the antisymmetrization concerns only
indices $[\mu^1]_{s_1}, \mu^2_{l_2} $ in
$[[\mu^1]_{s_1},\underbrace{\mu^2_1...\mu^2_{l_2-1}}\mu^2_{l_2}]$.

Combined description of all integer spin mixed-antisymmetric $ISO(1,d-1)$ group irreps can be
reformulated  with help of an auxiliary Fock space
$\mathcal{H}^f$, generated by $2$ pairs of fermionic creation
$a^i_{\mu^i}(x)$ and annihilation $a^{j+}_{\nu^j}(x)$ operators (in antisymmetric basis),
$i,j =1,2, \mu^i,\nu^j =0,1...,d-1$: ${a^i_{\mu^i},
a_{\nu^j}^{j+}}=-\eta_{\mu^i\nu^j}\delta^{ij}$
 and a set of constraints for an arbitrary string-like  vector
$|\Phi\rangle \in \mathcal{H}^f$,
\begin{eqnarray}
\label{PhysState}  \hspace{-1ex}&& \hspace{-2ex} |\Phi\rangle  =
\sum_{s_1=0}^{[d/2]}\sum_{s_2=0}^{s_1}
\Phi_{[\mu^1]_{s_1},[\mu^2]_{s_2}}(x)\,
\prod_{i=1}^2\prod_{l_i=1}^{s_i} a^{+\mu^i_{l_i}}_i|0\rangle,\\
\label{l0} \hspace{-1ex}&& \hspace{-3ex} \bigl({{l}}_0, {l}^i, l^{12},
t^{i_1j_1} \bigr)|\Phi\rangle = 0 ,
\quad l_0 = \partial^\mu\partial_\mu,\\
\label{lilijt} \hspace{-1ex} && \hspace{-2ex}  \bigl( {l}^i, l^{12},
t^{i_1j_1} \bigr) = \bigl(-i {a}^i_\mu \partial^\mu,
\textstyle\frac{1}{2}{a}^{1}_\mu {a}^{2\mu}, {a}^{1+}_\mu
a^{2\mu}\bigr)  .
\end{eqnarray}
 The set of $3$ even and $2$ odd, ${l}^i$,
primary constraints (\ref{l0}), (\ref{lilijt}) with $\{o_\alpha\}$
= $\bigl\{{{l}}_0, {l}^i, l^{12}, t^{12} \bigr\}$, because of
the property of translational invariance of the vacuum,
$\partial_\mu |0\rangle = 0$, are equivalent to
(\ref{Eq-0b})--(\ref{Eq-3b}) for all possible heights $s_1\geq s_2$. In turn, when we impose
on $|\Phi\rangle$  the additional to  (\ref{l0}), (\ref{lilijt})
constraints with number particles operators, $g_0^i$,
\begin{eqnarray}\label{g0iphys}
g_0^i|\Phi\rangle =(s_i-\textstyle\frac{d}{2}) |\Phi\rangle, \
 g_0^i = -\frac{1}{2}[{a}^{i+}_\mu,  {a}^{\mu{}i}],
\end{eqnarray}
these combined conditions are equivalent to Eqs.
(\ref{Eq-0b})--(\ref{Eq-3b}) for the field
$\Phi_{[\mu^1]_{s_1},[\mu^2]_{s_2}}(x)$ with
given spin $\mathbf{s} = (2, 2,  ... , 2,1,...,1)$.

The procedure of LF construction implies  the property of
BFV-BRST operator $Q$, $Q = C^\alpha o_\alpha + more$,  to be
Hermitian, that is equivalent to the requirements: $\{o_\alpha\}^+
= \{o_\alpha\}$ and closedness for $\{o_\alpha\}$ with respect to
the supercommutator multiplication $[\ ,\ \}$. Evidently,  the
set of $\{o_\alpha\}$ violates above conditions. To provide them
we consider in standard manner  an scalar product on
$\mathcal{H}^f$,
\begin{eqnarray}
\label{sproduct} \hspace{-1ex} && \hspace{-2ex} \langle{\Psi}|\Phi\rangle  =   \int
d^dx\sum_{s_1=0}^{[d/2]}\sum_{s_2=0}^{s_1}\sum_{p_1=0}^{[d/2]}\sum_{p_2=0}^{p_1}
\langle 0|\Bigl(\hspace{-1em}\prod_{(j,m_j)=(1,1)}^{2,p_j}
\hspace{-1em}{a}^{\nu^j_{m_j}+}_j\Bigr)^+\nonumber\\
\hspace{-1ex} && \hspace{-2ex} \times\Psi^*_{[\nu^1]_{p_1},[\nu^2]_{p_2}}
\Phi_{[\mu^1]_{s_1},[\mu^2]_{s_2}}\,
\prod_{(i,l_i)=(1,1)}^{(2,s_i)} {a}^{+\mu^i_{l_i}}_i|0\rangle .
\end{eqnarray}
As the result, the set of $\{o_\alpha\}$ extended by means of the
operators,
\begin{eqnarray} \label{lilijt+} \hspace{-2ex} && \hspace{-2.5ex} \bigl({l}^{i+},
l^{12+}, t^{12+} \bigr)  = \bigl(-i {a}^{i+}_\mu
\partial^\mu, \textstyle\frac{1}{2}{a}^{2+}_\mu {a}^{1\mu+},
{a}^{2+}_\mu {a}^{1\mu}\bigr) ,
\end{eqnarray}
is closed with respect to Hermitian conjugation, with taken into account of  $(l_0^+,\ {g_0^i}^+) = (l_0,\ {g_0^i})$. It is rather
simple exercise  to see the second requirement is fulfilled as
well if the number particles operators $g_0^i$ will be included
into set of all constraints $o_I$ having therefore the structure,
\begin{eqnarray}
\{o_I\} = \{o_\alpha, o_\alpha^+;\ g_0^i\}\equiv \{o_a, o_a^+ ;\
l_0,\ l^i,\ l^{i+};\ g_0^i\}. \label{inconstraints}
\end{eqnarray}
Together the sets $\{o_a, o_a^+\}$ in the Eq.
(\ref{inconstraints}), for $\{o_a\} = \{l^{12}, t^{12}\}$ and
 $\{o_A\}= \{l_0,\ l^i,\ l^{i+}\}$, may be considered from
the Hamiltonian analysis of the dynamical systems
 as the operator respective $4$ second-class and $5$ first-class constraints subsystems among
$\{o_I\}$ for topological gauge system (i.e. with zero Hamiltonian) because of,
\begin{eqnarray}
[o_a,\; o_b^+\} = f^c_{ab} o_c +\Delta_{ab}(g_0^i),\ [o_I,\;o_B\} =
f^C_{IB}o_C.
\label{inconstraintsd}
\end{eqnarray}
Here the constants $f^c_{ab}, f^C_{IB}$ are the antisymmetric with
respect to permutations of lower indices  and
quantities $\Delta_{ab}(g_0^i)$ form the non-degenerate $4\times
4$ matrix $\|{\mathrm{antidiag}}(-\Delta_{ab},\Delta_{ab})\|$ in the Fock space  $\mathcal{H}^f$ on
the surface $\Sigma \subset \mathcal{H}^f$:
$\|\Delta_{ab}\|_{|\Sigma} \ne 0 $, determined by the
equations, $(o_a, l_0,\ l^i)|\Phi\rangle = 0$.

Explicitly, operators $o_I$ satisfy to the Lie-algebra commutation
relations, $[o_I,\ o_J]= f^K_{IJ}o_K, \  f^K_{IJ}= - (-1)^{\varepsilon(o_I)\varepsilon(o_J)} f^K_{JI}$
with the structure constants $f^K_{IJ}$ being used in the
Eq.(\ref{inconstraintsd}),
%included the constants
%$f^{[g_0^i]}_{ab}: f^{[g_0^i]}_{ab}g_0^i \equiv
%\Delta_{ab}^{[g_0^i]}(g_0^i)$ there
and determined from the multiplication table~\ref{table in}.
\end{multicols} \hspace{-1ex}{\begin{table}
{{\footnotesize
\begin{center}
\begin{tabular}{||c||c|c|c|c|c|c|c||c||}\hline\hline
$\hspace{-0.2em}[\; \downarrow, \rightarrow
\}\hspace{-0.5em}$\hspace{-0.7em}&
 $t^{12}$ & $t^+_{12}$ &
$l_0$ & $l^i$ &$l^{i{}+}$ & $l^{12}$ &$l^{12{}+}$ &
$g^i_0$ \\
\hline\hline $t^{12}$
    & $0$ & $g_0^1-g_0^2$
   & $0$&\hspace{-0.3em}
    $\hspace{-0.2em}l^{2}\delta^{i{}1}$\hspace{-0.5em} &
    \hspace{-0.3em}
    $-l^{1+}\delta^{2{}i}$\hspace{-0.3em}
    &\hspace{-0.7em} $\hspace{-0.7em}0
    \hspace{-0.9em}$ \hspace{-1.2em}& \hspace{-1.2em}$
    0\hspace{-0.9em}$\hspace{-1.2em}& $-F^{12,i}$ \\
\hline $t^+_{12}$
    & $g_0^2-g_0^1$ & $0$
&$0$   & \hspace{-0.3em}
    $\hspace{-0.2em} l_{1}\delta^{i{}2}$\hspace{-0.5em} &
    \hspace{-0.3em}
    $-l^+_{2}\delta^{i{}1}$\hspace{-0.3em}
    & $0$ & $0$ & $F^{12,i+}$\\
\hline $l_0$
    & $0$ & $0$
& $0$   &
    $0$\hspace{-0.5em} & \hspace{-0.3em}
    $0$\hspace{-0.3em}
    & $0$ & $0$ & $0$ \\
\hline $l^j$
   & \hspace{-0.5em}$- l^{2}\delta^{j{}1}$ \hspace{-0.5em} &
   \hspace{-0.5em}$-
   l_{1}\delta^{j{}2}$ \hspace{-0.9em}  & \hspace{-0.3em}$0$ \hspace{-0.3em} & $0$&
   \hspace{-0.3em}
   $l_0\delta^{ji}$\hspace{-0.3em}
    & $0$ & \hspace{-0.5em}$ \textstyle\frac{1}{2}l^{[2+}\delta^{1]j}$
    \hspace{-0.9em}&$l^j\delta^{ij}$  \\
\hline $l^{j+}$ & \hspace{-0.5em}$l^{1+}
   \delta^{j{}2}$\hspace{-0.7em} & \hspace{-0.7em}
   $l_{2}^+\delta^{j{}1}$ \hspace{-1.0em} &
   $0$&\hspace{-0.3em}
      \hspace{-0.3em}
   $l_0\delta^{ji}$\hspace{-0.3em}
    \hspace{-0.3em}
   &\hspace{-0.5em} $0$\hspace{-0.5em}
    &\hspace{-0.7em} $ \textstyle\frac{1}{2}l^{[1}\delta^{2]j}
    $\hspace{-0.7em} & $0$ &$-l^{j+}\delta^{ij}$  \\
\hline $l^{12}$
    & \hspace{-0.3em}$0$
    \hspace{-0.5em} &\hspace{-0.5em} $0$\hspace{-0.3em}
   & $0$&\hspace{-0.3em}
    $0$\hspace{-0.5em} & \hspace{-0.3em}
    $ \hspace{-0.7em}\textstyle\frac{1}{2}l^{[2}\delta^{1]i}
    \hspace{-0.5em}$\hspace{-0.3em}
    & $0$ & \hspace{-0.7em}$ -\textstyle\frac{1}{4}(g_0^1+g_0^2)$\hspace{-0.7em}& $\hspace{-0.7em}  l^{12}\hspace{-0.7em}$\hspace{-0.7em} \\
\hline $l^{12+}$
    & $ 0$ & $ 0$
   & $0$&\hspace{-0.3em}
    $\hspace{-0.2em} \textstyle\frac{1}{2}l^{[1+}\delta^{2]i}$\hspace{-0.5em} & \hspace{-0.3em}
    $0$\hspace{-0.3em}
    & $\textstyle\frac{1}{4}(g_0^1+g_0^2)$ & $0$ &$\hspace{-0.5em}  -l^{12+}\hspace{-0.3em}$\hspace{-0.2em} \\
\hline\hline $g^j_0$
    & $F^{12,j}$ & $-F^{12}{}^{j+}$
   &$0$& \hspace{-0.3em}
    $\hspace{-0.2em}-l^i\delta^{ij}$\hspace{-0.5em} & \hspace{-0.3em}
    $l^{i+}\delta^{ij}$\hspace{-0.3em}
    & \hspace{-0.7em}$  -l^{12}$\hspace{-0.7em} & $ l^{12+}$&$0$ \\
   \hline\hline
\end{tabular}
\end{center}}} \vspace{-2ex}\caption{HS symmetry  superalgebra  $\mathcal{A}(Y[2],
\mathbb{R}^{1,d-1})$.\label{table in} }\end{table}
\vspace{-2ex}\begin{multicols}{2}
Note, that in the table~\ref{table in},  the squared brackets for the indices $i$,
$j$ in the quantity $A^{[i}B^{j]k}$
mean the antisymmetrization
$A^{[i}B^{j]k}$ = $A^{i}B^{jk}- A^{j}B^{ik}$ and $F^{12,i} =
   t^{12}(\delta^{i{}1}-\delta^{i{}2}),\quad F^{12,i+} =
   t^{12+}(\delta^{i{}1}-\delta^{i{}2})$.
We call the superalgebra of the operators $o_I$
as \emph{integer higher-spin symmetry algebra in Minkowski
space with a YT having $2$ columns} and denote it as $\mathcal{A}(Y[2], \mathbb{R}^{1,d-1})$.

The structure of $\mathcal{A}(Y[2], \mathbb{R}^{1,d-1})$  appears by insufficient to
 construct BRST operator $Q$ with  respect to its elements $o_I$ which should generate correct Lagrangian dynamics
 due to second-class constraints $\{o_a\}$ presence in it. Therefore, we should to convert $o_I$
 into enlarged set of operators $O_I$ with only first-class constraints.

%%%%%%%%%%%%%%%%%%%%%%%%%%%%%%%%
\section{Deformed  HS symmetry superalgebra for YT with $2$ columns}\label{Vermamodule}

We apply  an additive conversion procedure developed within BRST
method, (see e.g. \cite{BurdikPashnev}),   implying  the
enlarging of $o_I$ to $O_I = o_I + o'_I$, with additional parts
$o'_I$ supercommuting with all $o_I$ and determined on a new Fock space $\mathcal{H}'$. Now, the
elements $O_I$ are given on $\mathcal{H}^f\otimes \mathcal{H}'$ so
that a condition for $O_I$, $[O_I,\ O_J] \sim O_K$, leads to the
same algebraic relations for $O_I$ and $o'_I$ as those for $o_I$.

Because of only the generators which do not contain space-timer derivatives, $\partial_{\mu}$, are the second-class
constraints  in $\mathcal{A}(Y[2],\mathbb{R}^{1,d-1})$, i.e. $\{o'_a, {o'}^+_a\}$. Therefore,
one should to get new operator realization of  this  subalgebra. Note,  this subalgebra is isomorphic to $sl(2)\oplus sl(2)$.

An auxiliary oscillator realization of $sl(2)\oplus sl(2)$  algebra can be found by using Verma module concept \cite{Dixmier} and explicitly derived in
 the form
\begin{eqnarray}
&& t^{+\prime}_{12} \ = \ b_2^{+}, \quad {l_{12}^{+}}'\ = \ b_1^{+}, \nonumber \\
&& g_{0}^{i\prime} \ = \ h_1 + b_1^{+} b_1 +(-1)^i b_2^{+} b_2,\nonumber  \\
&& l_{12}' \ = \ -\textstyle \frac{1}{4} ( h_1 + h_{2 } + b_1^{+} b_1 ) b_1, \label{osc-}  \\
&& t_{12}' \ = \ -( h_2 - h_1  + b_2^{+} b_2 ) b_2,\nonumber
\end{eqnarray}
with new 2 pairs of bosonic creation (annihilation) operators $b^+_i(b_i)$, with non-trivial commutation relations, $[b_i,b_j^+]=\delta_{ij}$.
The operators $t^{+\prime}_{12}$ and $t^{\prime}_{12}$; $l^{+\prime}_{12}$ and $l^{\prime}_{12}$ are respectively Hermitian conjugated to each other, as well as  the number particles operators $g_{0}^{i\prime}$ is Hermitian with help of the Grassmann-even operator  $(K')^+=K'$ which should be found from the system of 4 equations,
\begin{eqnarray}
&& \langle{\Psi}|K't(l)_{12}'|\Phi\rangle\ =\  \langle{\Phi}|K't(l)_{12}^{+\prime}|\Psi\rangle^*, \nonumber\\
&& \langle{\Psi}|K'g_{0}^{i\prime}|\Phi\rangle\ =\  \langle{\Phi}|K'g_{0}^{i\prime}|\Psi\rangle^*.\label{systemK}
\end{eqnarray}
whose solution  may be presented in the form,
\begin{eqnarray}
%\label{explicit K}
 && \hspace{-2em} K' =  \sum_{n_{i} = 0}^{\infty}
 \frac{(-1)^{n_1+n_2}C_{h_1+h_2}(n_1)C_{h_2-h_1}(n_2)}{4^{n_1}n_{1}!n_{2}!(h_1+h_2+n_1)(h_2-h_1+n_2) } \nonumber \\
 && \hspace{-2em}  \label{Chn}\times|n_1,n_2\rangle \langle n_1,n_2 |,\texttt{ for  } C_{h}(n) = \prod_{i=0}^n (h+i),\
\end{eqnarray}
and $|n_1,n_2\rangle  =  (b_1^+)^{n_1}(b_2^+)^{n_2}|0\rangle.$
%%%%%%%%%%%%%%%%%%%%%%%%%%%%%%%%
\section{BRST-BFV operator and  Lagrangian formulations}\label{BRSToperator}
%%%%%%%%%%%%%%%%%%%%%%%%%%%%%%%%
Because of  algebra of $O_I$ under
consideration is a Lie superalgebra
$\mathcal{A}(Y[2],\mathbb{R}^{1,d-1})$ the BRST-BFV operator $Q'$
is constructed in the standard way
\begin{equation}\label{generalQ'}
    Q'  = {O}_I\mathcal{C}^I + \textstyle\frac{1}{2}
    \mathcal{C}^I\mathcal{C}^Jf^K_{JI}\mathcal{P}_K(-1)^{\epsilon({O}_K) + \epsilon({O}_I)}
\end{equation}
with the constants $f^K_{JI}$ from the table~\ref{table in},
constraints $O_I = (l_0$, $l^+_i$, $l_i$;   $L_{12}, L^+_{12},
T_{12}$, $T^+_{12}, G_0^i)$, fermionic [bosonic]  ghost fields and
conjugated to them momenta $(C^I, \mathcal{P}_I)$  =
$\bigl((\eta_0, {\cal{}P}_0);
 (\eta_{12}, {\cal{}P}^+_{12})$;
$(\eta^+_{12}, {\cal{}P}_{12j}); (\vartheta_{12},\lambda^+_{12})$;
$(\vartheta^+_{12}, \lambda_{12}); (\eta^i_{G},
{\cal{}P}^i_{G})$; $[ (q_i, p_i^+), (q_i^+, p_i)]\bigr)$  with non-vanishing (anti)commutators
\begin{eqnarray}
 \hspace{-1em}&&\hspace{-1em}
  \{\vartheta_{12},\lambda^+_{12}\}= \{\eta_{12},{\cal{}P}_{12}^+\} = 1,\  [q_i, p^{+}_j] = \delta_{ij}
\end{eqnarray}
 and for zero-mode ghosts  $\{\eta_0,{\cal{}P}_0\}= \imath$,
$\{\eta^i_{\mathcal{G}}, {\cal{}P}^j_{\mathcal{G}}\}
 = \imath\delta^{ij}$. The ghosts possess the
 standard  ghost number distribution,
$gh(\mathcal{C}^I)$ = $ - gh(\mathcal{P}_I)$ = $1$
$\Longrightarrow$  $gh({Q}')$ = $1$. Therefore, BRST-BFV operator $Q'$ and $Q$ are determined as
\begin{eqnarray}\label{Q'}
\hspace{-1em}&&\hspace{-0.7em} Q'  =   Q +
\eta^i_{G}(\sigma^i+h^i)+\mathcal{B}^i \mathcal{P}^i_{G},\quad
\texttt{with some $\mathcal{B}^i$,}\\
\label{Q} \hspace{-1em}&&\hspace{-0.7em} Q = \eta_0 L_0\hspace{-0.1em} +\hspace{-0.1em} i  q_iq_i^+  P_0
 \hspace{-0.1em} + \hspace{-0.1em} \Delta Q, \  \Delta Q  \hspace{-0.1em}=\hspace{-0.1em}  \Bigr( q_il^+_i \hspace{-0.1em} + \hspace{-0.1em} \eta_{12}L^+_{12}      \\
 \hspace{-1em}&&\hspace{-0.7em}  + \vartheta_{12} T^+_{12} + \textstyle\frac{1}{2}\epsilon_{ij}\eta_{12} q_i^+ p_j^+
  + \vartheta_{12}(q_2^+p_1+q_1p_2^+ ) +h.c.\Bigr)  \nonumber\\
\hspace{-1em}&&\hspace{-0.7em}
  \sigma_i+h_i=  G_0^{i} - q_i^+ p_i - q_i p_i^+   +\eta_{12}^+ \mathcal{P}_{12} -\eta_{12} \mathcal{P}^+_{12} \nonumber\\
\hspace{-1em}&&\hspace{-0.7em}\ +(-1)^i( \vartheta_{12}^+\lambda_{12}-\vartheta_{12}\lambda_{12}^+ )
\label{separation2} ,
\end{eqnarray}
with real $\epsilon_{ij}=-\epsilon_{ji}, \epsilon_{12}=1$. The property of $Q'$ to be Hermitian in $\mathcal{H}_{tot}$, $\mathcal{H}_{tot}$ = $\mathcal{H}^f \otimes \mathcal{H}'$ $\otimes \mathcal{H}_{gh}$ is determined by the rule
\begin{eqnarray}\label{HermQ}
Q^{\prime +}K\  =\ K Q'\texttt{  where   } K\ = \ 1 \otimes K' \otimes 1_{gh}.
\end{eqnarray}
To construct Lagrangian formulation for bosonic  HS fields subject to $Y[s_1,s_2]$ we choose a representation of
$\mathcal{H}_{tot}$: $(q_i, p_i,  \eta_{12}$, $\vartheta_{12},
\mathcal{P}_0$, $ \mathcal{P}_{12}$, $\lambda_{12},
\mathcal{P}^{i}_G)|0\rangle=0$, and suppose that the field vectors
$|\chi \rangle$, as well as the gauge parameters $|\Lambda \rangle$
do not depend on not constrained ghosts $\eta^{i}_G$, 
extend our basic vector  $|\Phi\rangle$ (\ref{PhysState}) given in $\mathcal{H}^f$  to
\vspace{-1ex}\begin{eqnarray}
\hspace{-1em}&&\hspace{-0.7em} |\chi\rangle =
\sum_{\{n\}_b=0}^{\infty}
\sum_{\{n\}_f=0}^{1}
\eta_{0}^{n_{\eta_{0}}}
\eta_{12}^{+n_{\eta_{12}}}
\vartheta_{12}^{+n_{\vartheta_{12}}}
\mathcal{P}_{12}^{+n_{P_{12}}}
\lambda_{12}^{+n_{\lambda_{12}}} \nonumber
\\
\hspace{-1em}&&\hspace{-0.7em} \times
\prod_{i=1}^{2}
(\eta_{i}^{G})^{ n_{i}}
q_i^{+n_{q_i}}
p_i^{+n_{p_i}}
b_i^{+n_{b_i}}
\left|\Phi({a_i}^+)_{\{n\}_f \{n\}_b}\rangle\right. ,
\label{extState}
\end{eqnarray}
where the integers
%\begin{eqnarray}
$\{n\}_b = n_{q_i},n_{p_i},n_{b_i} \in \mathbb{N}$ and $\{n\}_f$ = $n_{\eta_0}$, $n_{\eta_{12}},n_{P_{12}},n_{\vartheta_{12}},n_{\lambda_{12}} \in \mathbb{Z}_2$.

From the BRST-like equation, determining the
physical vector (\ref{extState}) and from the set of reducible gauge
transformations, homogeneous in ghost number
 ${Q}'|\chi^0\rangle = 0$ and 
$\delta|\chi\rangle$ = $Q'|\Lambda^0\rangle$, $\delta|\Lambda^0\rangle =
Q'|\Lambda^1\rangle$, $\ldots$, $\delta|\Lambda^{(r-1)}\rangle =
Q'|\Lambda^{(r)}\rangle$, for $gh(|\chi\rangle)=gh(|\Lambda^{(k)}\rangle)+k+1=0$, the decomposition in $\eta^{i}_G$ leads to the relations:
\begin{eqnarray}
\label{Qchi} \hspace{-1em}&&\hspace{-0.5em}  \bigl(Q|\chi^0\rangle, \delta|\chi^0\rangle, ..., \delta|\Lambda^{(r-1)}\rangle\bigr) =\bigl(0, Q|\Lambda^0\rangle, ..., Q|\Lambda^{(r)}\rangle\bigr),\nonumber \\
\hspace{-1em}&&\hspace{-0.5em}  [\sigma^i+h^i]\bigl(|\chi^0\rangle, |\Lambda^0\rangle, \ldots ,
|\Lambda^{(r)}\rangle\bigr) = 0, \label{QLambda}
\end{eqnarray}
with  $r =
s_1 + s_2$  being  the stage of reducibility both for
massless and for the massive bosonic HS field.
Resolution  the
spectral  problem from (\ref{QLambda})
yields   the eigenvectors of
the operators $\sigma^i$: $|\chi^0\rangle_{[n]_2}$,
$|\Lambda^0\rangle_{[n]_2}$, $\ldots$, $|\Lambda^{r}\rangle_{[n]_2}$,
for $[n]_2 = [n_1,n_2]$, $n_1 \geq n_2 \geq 0$ and corresponding
eigenvalues of the parameters $h^i$ (for massless  HS fields  $i=1,2$),
\begin{eqnarray}
\label{hi} \hspace{-1em}&&\hspace{-1.5em}  -h^i = n_i - \frac{d}{2}  -(-1)^i \;, \
 n_1, \in \mathbb{Z}, n_2 \in
\mathbb{N}_0\,.
\end{eqnarray}
 One can show, first, the operator $Q$ is nilpotent on the subspaces determined by the solution for (\ref{QLambda}), second,    to construct Lagrangian for the field corresponding to a definite YT
(\ref{Young k}) we must put  $n_i=s_i$, and, third,  one should substitute
$h^i$ corresponding to the chosen $n_i$ (\ref{hi})  into  $Q$
(\ref{Q'}) and relations (\ref{QLambda}).
 Thus, the equation of
motion (\ref{Qchi}) corresponding to the field with a given
$Y[s_1,s_2]$ has the form
\begin{eqnarray}
Q_{[s]_2}|\chi^0\rangle_{[s]_2}=0, \ \mathrm{for}\  |\chi^0\rangle_{[s]_2}{}_{(\{n\}_f= \{n\}_b =0)} =  |\Phi\rangle_{[s]_2} \label{Q12}
\end{eqnarray}
Because of commutativity $[Q,\sigma_i\}=0$ we have joint system of proper eigen-functions $|\chi^l\rangle_{[s_1,s_2]}$ for $l=0,1,..., s_1+s_2+1$ and  eigen-values $h^i(s_i)$ so that
the sequence of reducible gauge transformations for the field with
given $[s_1,s_2]$ are described (for $k=1,...,\sum_{i=1}^2 s_i$) by:
\begin{eqnarray}
\label{dx0} \hspace{-1em}&&\hspace{-1.5em}\delta|\chi^0 \rangle_{[s]_2}
=Q_{[s]_2}|\Lambda^{(0)}\rangle_{[s]_2}, \
\delta|\Lambda^{(0)}\rangle_{[s]_2} =Q_{[s]_2}|\Lambda^{(1)}\rangle_{[s]_2}
\,,
\nonumber \\
\label{dxs} \hspace{-1em}&&\hspace{-1.5em}  \delta|\Lambda^{(k-1)} \rangle_{[s]_2}
=Q_{[s]_2}|\Lambda^{(k)}\rangle_{[s]_2}, \
\delta|\Lambda^{(s_1+s_2)} \rangle_{[s]_2} =0.
\end{eqnarray}
The  equation of
motion (\ref{Q12}) are Lagrangian with appropriate numbers of auxiliary HS fields and  derived from a gauge-invariant  Lagrangian action (for $ K_{[s]_2}=K\vert_{h^i =h^i(s)}$)
\begin{eqnarray} {\cal
S}_{[s]_2} = \int d \eta_0  {}_{[s]_2}\langle \chi^0 |K_{[s]_2}
Q_{[s]_2}| \chi^0 \rangle_{[s]_2}. \label{S}
\end{eqnarray}

%%%%%%%%%%%%%%%%%%%%%%%%%%%%%%%%%%%%%%%%%%%%%%%%%%%%%%%%%%%%%%%%
\section{Constrained Lagrangian Formulations}\label{constr}
%%%%%%%%%%%%%%%%%%%%%%%%%%%%%%%%%%%%%%%%%%%%%%%%%%%%%%%%%%%%%%%%
Let us list the key points of the derivation of the constrained LF from unconstrained one for the same bosonic field subject to $Y[s_1,s_2]$
\begin{enumerate}
    \item reduction of HS symmetry algebra $\mathcal{A}(Y[2],\mathbb{R}^{1,d-1})$ $\to$
    $\mathcal{A}_{r}(Y[2],\mathbb{R}^{1,d-1}) =
    \frac{\mathcal{A}(Y(k),\mathbb{R}^{1,d-1})}{sl(2)\oplus sl(2}$ =
    $\{l_0,l_i,l_j^+\}$;
    \item absence of the  2nd class constraints for ($m=0$) $\Longrightarrow$
    absence of the conversion procedure;
    \item  reduction of $Q'$
    (\ref{Q'}) to\\ {$Q_r$} = $\eta_0 l_0
    + \sum_i(q_il_i^++ q_i^+l_i + \imath
    q_i^+q^i{\cal{}P}_0)$;
\item presence of $2$ off-shell BRST extended by
$q_i, q_i^+, {p}^i, {p}_i^+$ constraints
$\mathcal{L}_{12},
\mathcal{T}_{12}$, and spin operator
\begin{equation}
\hspace{-0.5em}{\sigma^i_r} = g_0^i
+ q_i p^+_i +  q_i^+ p_i:\  [\mathcal{A},Q_r] =0,
\end{equation}
for $\mathcal{A}\in\{\mathcal{L}_{12},\mathcal{T}_{12},\sigma^i_r\}$ which look  explicitly as
\begin{eqnarray}
\hspace{-1em}\mathcal{L}_{12} =  {l}_{12}+  \textstyle\frac{1}{2}\epsilon_{ij} q_i p_j, \  \mathcal{T}_{12}  =  t_{12} +   q_2p_1^++q_1^+p_2. \label{eontr}
\end{eqnarray}
\end{enumerate}
The proper constrained Lagrangian action  is determined by the relations (with obvious BRST-complex of reducible gauge transformations generated by $Q_r$)
\begin{eqnarray}
\hspace{-0.0em}\mathcal{S}_{r[s]_2}\hspace{-0.1em} = \hspace{-0.1em}\int\hspace{-0.1em} d \eta_0  {}_{[s]_2}\langle \chi^0_r
|Q_r| \chi^0_r \rangle_{[s]_2}, \; (\mathcal{L}_{12}, \mathcal{T}_{12})| \chi^k_r \rangle=0. \label{Sclc}
\end{eqnarray}
\section{Conclusion}
%%%%%%%%%%%%%%%%%%%%%%%%%%%%%%%%
Thus, we have constructed  gauge-invariant unconstrained and constrained
Lagrangian descriptions of free integer  HS fields belonging to an
irreducible representation of the Poincare group $ISO(1,d-1)$ with
the arbitrary Young tableaux having $2$ columns in the ``metric-like"
formulation. The results of this study are the general and
obtained on the base of universal method which is applied by the
unique way to both massive and massless bosonic HS fields with a
mixed antisymmetry in a Minkowski space of any dimension.
%%%%%%%%%%%%%%%%%%%%%%%%%%%%%%%%
\section*{Acknowledgement}
This research has been supported by the  RFBR grant, project
No. 12-02-00121 and by the grant for LRSS, project No.  88.2014.2.
\end{multicols}
%%%%%%%%%%%%%%%%%%%%%%%%%%%%%%%%

\end{document}